# Tracking the Twitter attention around the research efforts on the COVID-19 pandemic

Zhichao Fang[1*], Rodrigo Costas[1,2]

* Corresponding author

Zhichao Fang

[1] Centre for Science and Technology Studies (CWTS), Leiden University, Leiden, The Netherlands.

E-mail: z.fang@cwts.leidenuniv.nl

ORCID: 0000-0002-3802-2227

Rodrigo Costas

[1] Centre for Science and Technology Studies (CWTS), Leiden University, Leiden, The Netherlands.

[2] DST-NRF Centre of Excellence in Scientometrics and Science, Technology and Innovation Policy, Stellenbosch University, Stellenbosch, South Africa.

E-mail: rcostas@cwts.leidenuniv.nl

ORCID: 0000-0002-7465-6462

## Abstract

The outbreak of the COVID-19 pandemic has been accompanied by a bulk of scientific research and related Twitter discussions. To unravel the public concerns about the COVID-19 crisis reflected in the science-based Twitter conversations, this study tracked the Twitter attention around the COVID-19 research efforts during the first three months of 2020. On the basis of nearly 1.4 million Twitter mentions of 6,162 COVID-19-related scientific publications, we investigated the temporal tweeting dynamic and the Twitter users involved in the online discussions around COVID-19-related research. The results show that the quantity of Twitter mentions of COVID-19-related publications was on rising. Scholarly-oriented Twitter users played an influential role in disseminating research outputs on COVID-19, with their tweets being frequently retweeted. Over time, a change in the focus of the Twitter discussions can be observed, from the initial attention to virological and clinical research to more practical topics, such as the potential treatments, the countermeasures by the governments, the healthcare measures, and the influences on the economy and society, in more recent times.

## Introduction

The global spread of the COVID-19 pandemic, an infectious disease caused by the pathogen severe acute respiratory syndrome coronavirus 2 (SARS-CoV-2), has already unleashed an unprecedented impact on public health, economy, and human society worldwide (McKee & Stuckler, 2020). As of June 2, 2020, it is reported by the World Health Organization (WHO) that there have been over 6.1 million confirmed cases of COVID-19 globally, carrying a mortality of approximately 6%.[1] On January 30, 2020, the WHO officially declared that the COVID-19 outbreak constitutes a Public Health Emergency of International Concern (PHEIC),[2] making it become the sixth PHEIC in the 21st century after the 2009 H1N1 pandemic, the 2014 Polio declaration, the 2014 Ebola virus disease (West Africa), the 2016 Zika virus epidemic, and the 2018 Ebola virus disease (Kivu) (Harmer et al., 2020).

[1] The numbers of confirmed cases and deaths of COVID-19 were retrieved from the WHO coronavirus disease (COVID-19) dashboard: https://covid19.who.int/. Accessed 2020-06-02.

In response to this ongoing public health emergency, scientists around the world have been contributing their expertise to the understanding and potential treatments of the novel coronavirus disease, leading to an explosion of research outputs covering a range of subject fields (Callaway et al., 2020). The strong concerns about this public health crisis arose within not only the scientific community but also the social media landscape. Millions of people are talking about the coronavirus on social media (Yammine, 2020), particularly on Twitter, where there are massive conversations around a variety of topics related to COVID-19 (Chen et al., 2020; Thelwall & Thelwall, 2020). Amongst these conversations, up-to-date research progress made by scientists is one of the most important elements, reflecting the Twitter attention paid toward scientific discoveries in fighting the pandemic. Against this background, through the lens of scholarly Twitter metrics, which focus on the recorded events of acts on Twitter related to scholarly documents or scholarly agents (Haustein, 2019), this study aims to disclose the focus and dynamic of public concerns about relevant research efforts during the time of PHEIC, specifically, in the case of the COVID-19 pandemic.

*Twitter attention toward scientific research*

As one of the most prevalent social media platforms, the role that Twitter plays in scholarly communication has been widely investigated in previous studies (Sugimoto et al., 2017). On the whole, Twitter mention data outperform most other altmetric indicators in terms of not only the data coverage but also the accumulation speed after publication (Costas et al., 2015; Fang & Costas, 2020; Haustein et al., 2015), making it possible to measure the Twitter reception of research outputs in a relatively short period of time.

Considering the above advantages, the potential of Twitter metrics in research evaluation has already been discussed from both conceptual and practical perspectives (Wouters et al., 2019). Moreover, captured Twitter attention toward scientific publications has been analyzed for identifying the focus of interest of Twitter users on some specific subject fields or research topics. For example, Robinson-Garcia et al. (2019) mapped Twitter attention distributing within the field of Microbiology and found that topics about translational medicine, future prospects and challenges, and bacterial outbreaks are prominent with regard to their Twitter mentions received. Haunschild et al. (2019) made comparisons for a set of climate change publications among the networks created based on the author keywords and the tweet hashtags, finding that most tweeted research topics are those about the consequences of climate change for humans. Taking the area of Big Data as a case study, Lyu and Costas (2020) observed that hashtags embedded in the Twitter mentions of big data research are mainly about technologies, showing a similar technical orientation as author keywords in publications. These existing studies indicate that detected Twitter attention toward scientific research opens a window for tracking broader public concerns beyond academia about specific topics. The study of Twitter activities around scholarly outputs can be used to analyze broader perspectives on science-society interactions in what has been labeled as "heterogeneous couplings" (Costas et al., 2017), these being relevant to characterize and study how in a pandemic scientific results are being received and communicated among very diverse audiences.

*Scientometric perspectives on COVID-19-related research*

The outbreak of COVID-19 is not merely an urgent threat to global health, but also a significant challenge to the current scientific system. On the one hand, being confronted with such emergency, deficiencies existing in the scholarly communication system have further been proven to be obstacles in the way to a more open and efficient scholarly environment (Larivière et al., 2020). On the other hand, the transformation from the traditional status of journals to faster online publishing channels has been facilitating the explosion of COVID-19-related scientific literature, resulting in an over-abundance of scientific information (Brainard, 2020; Torres-Salinas et al., 2020).

[2] See more details about the WHO director-general's statement at: https://www.who.int/dg/speeches/detail/who-director-general-s-statement-on-ihr-emergency-committee-on-novel-coronavirus-(2019-ncov). Accessed 2020-06-02.

To delineate the massive research progress, there are a range of studies investigating COVID-19-related literature from several scientometric perspectives, such as the coverage of publications in diverse scholarly databases (Kousha & Thelwall, 2020), the identification of hot topics (Haghani et al., 2020), the international collaboration patterns, and the funding sources (Fry et al., 2020; Zhang et al., 2020), etc. Amongst these existing studies, some of them have a focus on the social media performance of COVID-19-related publications. For instance, Colavizza et al. (2020) took relatively exhaustive investigations to the CORD-19 database, namely, the COVID-19 Open Research Database which captures global research on COVID-19 and coronavirus family of viruses (Wang et al., 2020), and concluded that the CORD-19 database covers a wide range of research on viruses in general but not only on COVID-19 and coronavirus. In particular, they found that CORD-19 publications published in 2020, especially those on topics of pressing relevance, are disproportionately popular on social media. Torres-Salinas et al. (2020) focused on the uptake of open access on COVID-19-related literature and found that nearly 67.5% of publications in their dataset are openly accessible, moreover, they confirmed the advantage of open access publications in obtaining social media attention over non open access publications. Based on the relationships between early citations (in the database Dimensions.ai) and social media mentions, Kousha and Thelwall (2020) observed a high degree of convergence between COVID-19-related publications shared in the social web and citation counts, suggesting that altmetrics, Twitter counts in particular, might be helpful for quickly filtering useful new documents from the daily flood of COVID-19-related literature.

*Objectives*

Taken together, although the coverage and performance of COVID-19-related publications in social media have been analyzed in many existing studies with diverse data sources. In contrast, there are less explorations into what and how the public discuss about COVID-19-related research progress on social media. In response, focusing on the Twitter attention toward COVID-19-related research during the time of PHEIC, the main objectives of this study are two-fold: the first one is to unravel the tweeting patterns around COVID-19-related research from the perspectives of the temporal accumulation and engaging Twitter users; and the second one is to trace the evolution of the focus of the Twitter discussions around COVID-19-related research over time. We addressed the following specific research questions:

RQ1. How are the COVID-19-related scientific publications mentioned on Twitter over time since the outbreak?

RQ2. By drawing on the Twitter users' profile descriptions, who are the actors tweeting COVID-19-related research? What kinds of Twitter users are more influential in terms of being retweeted?

RQ3. Based on the titles, what are the main research topics of COVID-19-related publications? Which research topics attract higher levels of Twitter attention?

RQ4. On the basis of the full tweet texts and hashtags, what are the topics raised by the Twitter users when discussing the COVID-19-related publications? How does the focus of interest of their Twitter discussions change over time?

## Data and methods

*Dataset of COVID-19-related publications*

Ever since the outbreak of COVID-19, an increasing number of academic publishers and organizations have been gathering and compiling global relevant research and making them open access for the sake of public concerns. Our meta-dataset stemmed from two databases: one is the database of COVID-19-related publications established and updated by the WHO,[3] the other one is the file made available by Dimensions.ai, which contains all COVID-19-related published articles, preprints, datasets and clinical trials from Dimensions.ai.[4] On April 4, 2020, a total of 8,163 distinct scientific publications were extracted from these two data sources, consisting of 3,478 publications tracked by the WHO and 6,338 publications provided by Dimensions.ai (1,653 publications are overlapped).

The bibliometric information of the above set of publications were retrieved through the Dimensions.ai API on April 4, 2020 as well. On the whole, 6,582 of them (accounting for 80.6%) are indexed by Dimensions.ai with the DOI or PubMed ID matched. Since we focus on the research efforts occurred in the first three months of 2020 (from January 1, 2020 to March 31, 2020), publications published beyond this time window were filtered out. In this study, DOI created date, namely, the date on which a DOI was created, was collected through the Crossref API to serve as the more precise proxy of publication date (Fang & Costas, 2018). While the DOI created date is unavailable, publication date recorded by Dimensions.ai was used as the alternative.[5] Based on the publication date, a total of 6,162 publications were selected as the final dataset.

*Twitter mention data of COVID-19-related publications*

In order to track Twitter attention toward COVID-19-related publications, Twitter mention data of selected publications recorded by Altmetric.com were queried through the API between April 4 and April 7, 2020. Based on the list of tweet IDs responded by the Altmetric.com API, we further collected detailed Twitter mention information with the Twitter API on April 7, 2020. Finally, after excluding unavailable Twitter mentions caused by the deletion of tweets and the suspension or protection of Twitter accounts (Fang et al., 2020), a total of 4,195 publications (accounting for 68.1%) were found to have some Twitter mention data accrued in the same time window as the publication date of selected publications, involving with 1,374,231 distinct tweets posted by 655,494 unique Twitter users.

On the basis of tweet objects information responded by the Twitter API,[6] specifically the retweeted, replied, and quoted relationships of each tweet, Twitter mentions were classified into two main categories: original tweet (including regular original tweet and reply tweet) and retweet (including simple retweet and quote tweet). Table 1 lists the concepts of these types of tweets, together with their presence in our dataset. Simple retweets contribute most of the Twitter mentions of COVID-19-related publications, accounting for 76.1%. Note that in this study, "original tweet" and "retweet" refer to the two main categories. Since except for simple retweet, the other three sub-categories contain original contents from the Twitter users, "tweets with original contents" analyzed in the text analysis parts of this study considered regular original tweets, reply tweets, and quote tweets.

---

[3] See more information about the COVID-19-related database of WHO at: https://www.who.int/emergencies/diseases/novel-coronavirus-2019/global-research-on-novel-coronavirus-2019-ncov. Accessed 2020-05-20.

[4] See more information about the file provided by Dimensions.ai at: https://www.dimensions.ai/news/dimensions-is-facilitating-access-to-covid-19-research/. Accessed 2020-05-20.

[5] Amongst 6,582 publications indexed by Dimensions.ai, 6,300 of them have a DOI created date recorded by Crossref as their publication date, while the remaining 282 publications were assigned with the Dimensions.ai publication date.

[6] See more introduction to tweet objects in the JSON data responded by the Twitter API at: https://developer.twitter.com/en/docs/tweets/data-dictionary/overview/tweet-object. Accessed 2020-05-20.

**Table 1**. Classification of tweets and corresponding concepts

| Main category | Sub-category | Concept and method of generating | Containing original contents | N | P |
|---|---|---|---|---|---|
| Original tweet | Regular original tweet | A tweet originally posted by a Twitter user. It is generated by typing in the compose box and clicking the Tweet button. | Yes | 149,431 | 10.9% |
| | Reply tweet | A reply tweet is a response to a tweet. It is generated by clicking the Reply icon from a tweet, typing in the compose box, and clicking the Reply button. | Yes | 101,078 | 7.3% |
| Retweet | Simple retweet | A simple retweet is a re-posting of a tweet by simply using Twitter's native retweet functionality without comments attached. It is generated by clicking the Retweet icon from a tweet and selecting "Retweet" directly. | No | 1,046,058 | 76.1% |
| | Quote tweet | A quote tweet is a re-posting of a tweet by using Twitter's native retweet functionality with original comments attached. It is generated by clicking the Retweet icon from a tweet and selecting "Retweet with comment" to add comments. | Yes | 77,664 | 5.7% |

Note: Table 1 only briefly introduces the methods of generating different types of tweets by manual operation on Twitter. It should be noted that tweets can be created programmatically by APIs as well. See more introduction to the concepts and methods of generating different categories of tweets at: https://help.twitter.com/en/using-twitter. (Accessed 2020-05-20). Moreover, because one can reply to a tweet with quoting other tweets in the reply tweet, it is possible for a tweet to be classified as a reply tweet and a quote tweet at the same time. Note that in this study, to avoid the overlap in the calculation of numbers of this kind of tweets with dual sub-categories, they were only counted as reply tweets because the following quoting behavior is resulted by the replying action happened at first.

In addition to the category assigned, detailed information responded by the Twitter API were also appended to each tweet for further analysis, including the full tweet text, the post date (i.e., the date on which the tweet was posted), the hashtags used, the number of retweets received, and the language of tweet text, etc., as well as the information of Twitter users, such as the profile descriptions.

*Analytical approaches*

To show overall pictures of the COVID-19 dataset and the Twitter discussions around them, VOSviewer (van Eck & Waltman, 2010) was utilized as the visualization tool for text mining and constructing co-occurrence networks of terms extracted from the titles of publications, the descriptions of Twitter users, and the tweet texts, respectively, as well as for creating *hashtag coupling*[7] networks. Besides, the so-called overlay visualization of VOSviewer was employed to exhibit additional information on top of the generated base maps, such as the average number of Twitter mentions received by terms, the average number of retweets received by Twitter users, and the post date of Twitter mentions and their appendant hashtags.

For measuring how recently the terms and hashtags in Twitter mentions were posted by users, the post month and post day of tweets were used to create the post date score, with post month serving as the integer part and post day the decimal part (e.g., the post date score of a tweet posted on February 1 is 2.01). The higher the post date score, the more recently the tweet was posted. The post date score of each tweet was also applied to its used terms and hashtags.

In order to examine the extent to which the tweet texts are different from the corresponding titles of publications, we cleaned the full texts of tweets with original contents in English at first by removing the embedded URLs, handle names of mentioned Twitter users and the @-sign, as well as the #-sign of hashtags. The texts of hashtags were kept because they often contain specific meaning. In this study the text similarity between titles of publications and related cleaned tweet texts was measured by cosine similarity score. Cosine similarity scores were calculated for each pair of texts by using the Python *scikit-learn* package (Pedregosa et al., 2011). The higher the cosine similarity score, the more similar the two pieces of texts.

[7] Once a publication is tweeted with two hashtags attached (even if in different tweets), there is a so-called coupling relationship between these two hashtags (Costas et al., 2017).

It should be noted that in the analyses involved with full tweet texts, we only took into consideration the 209,003 tweets with original contents in English language, which account for the majority (about 63.7%) in the sub-categories of all tweets with original contents (as depicted in Table 1). Tweet texts in other languages were not included in the text analysis part in this study.

## Results

Coinciding with the four proposed research questions, this section characterizes the Twitter attention toward COVID-19-related research from four perspectives. The first one studies the temporal accumulation patterns of tweets mentioning COVID-19-related research over time, with a focus on the top-10 most tweeted publications. The second perspective shows the profile characteristics of Twitter users who have participated in sharing COVID-19-related research to the Twitter landscape. The third perspective presents an overall picture of research topics of COVID-19-related publications, and identifies those with higher levels of Twitter attention received. The last perspective analyzes the discussions around COVID-19-related research on Twitter based on both full tweet texts and hashtags used, to disclose Twitter users' focus of interest and its evolution over time.

### *Exploring the temporal distribution of Twitter mentions of COVID-19-related research*

Since the reported outbreak of COVID-19, scientists from around the world have made remarkable contributions to the understanding of this novel infectious disease, which is represented by the increasing number of related publications. As shown in Figure 1(a), the quantity of COVID-19-related publications presents a rising tendency in the first three months of 2020, particularly in late March. Around 68.1% of COVID-19-related publications in our dataset have been mentioned by Twitter users at least once. Figure 1(b) shows the temporal distribution of the Twitter mentions. It is obvious that intensive Twitter attention toward COVID-19-related research started to rise on January 21, 2020, and reached the first peak on January 25, following the Wuhan city's lockdown on January 23. In February when the outbreak was temporarily confined to China, Twitter attention toward COVID-19-related research maintained a relatively stable state. However, along with the global outbreak in full swing and the mounting research outputs in March, Twitter attention steeply increased, especially after March 15 when the global confirmed cases outside of China exceeded those reported in China for the first time.

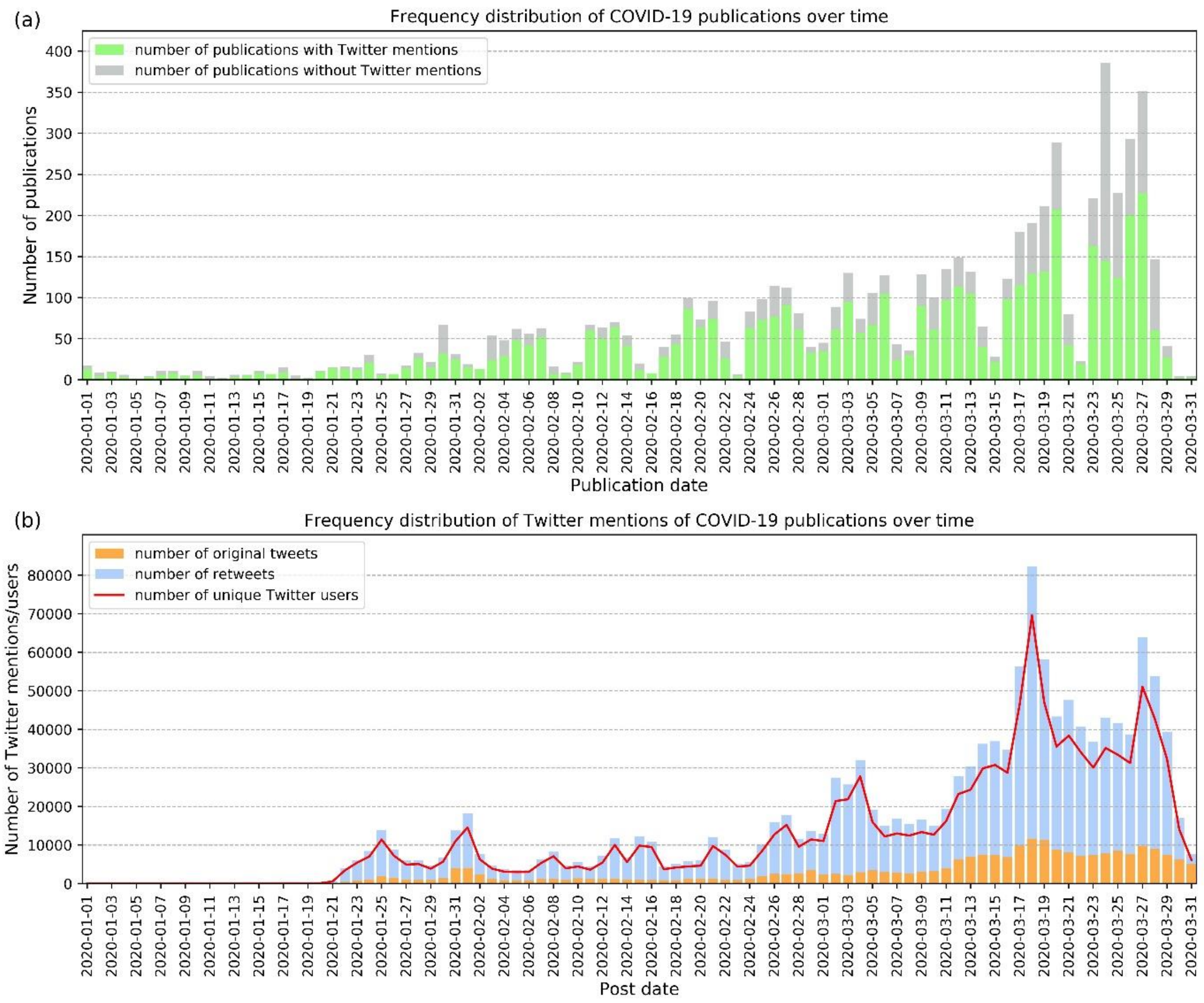

**Figure 1**. Temporal distribution of (a) COVID-19 publications and (b) related Twitter mentions.

To provide a more detailed analysis about how COVID-19-related publications were shared and discussed after they have been published, we selected the top-10 most tweeted publications from the dataset as a case study. Table 2 lists the information of these ten publications ranked by their number of Twitter mentions accrued. The publication entitled “The proximal origin of SARS-CoV-2”, whose main conclusion is that SARS-COV-2 is not a purposefully manipulated virus from laboratories (Andersen et al., 2020), attracted the most attention on Twitter in our observation time window, followed by research related to clinical treatments, public health countermeasures, clinical characteristics, and transmissibility of the virus, etc.

**TABLE 2.** Top-10 COVID-19-related publications with the most Twitter mentions

| Rank | Title | DOI | Journal/ Source | Publication date | N_tweets |
|---|---|---|---|---|---|
| #1 | The proximal origin of SARS-CoV-2. | 10.1038/s41591-020-0820-9 | Nature Medicine | 2020-03-17 | 70,881 |
| #2 | COVID-19 outbreak on the Diamond Princess cruise ship: estimating the epidemic potential and effectiveness of public health countermeasures. | 10.1093/jtm/taaa030 | Journal of Travel Medicine | 2020-02-28 | 29,077 |
| #3 | Clinical characteristics of coronavirus disease 2019 in China. | 10.1056/nejmoa2002032 | New England Journal of Medicine | 2020-02-28 | 24,822 |
| #4 | Uncanny similarity of unique inserts in the 2019-nCoV spike protein to HIV-1 gp120 and Gag. | 10.1101/2020.01.30.927871 | bioRxiv | 2020-01-31 | 21,360 |
| #5 | Aerosol and surface stability of SARS-CoV-2 as compared with SARS-CoV-1. | 10.1056/nejmc2004973 | New England Journal of Medicine | 2020-03-17 | 21,222 |
| #6 | Substantial undocumented infection facilitates the rapid dissemination of novel coronavirus (SARS-CoV2). | 10.1126/science.abb3221 | Science | 2020-03-16 | 21,152 |
| #7 | High temperature and high humidity reduce the transmission of COVID-19. | 10.2139/ssrn.3551767 | SSRN Electronic Journal | 2020-03-10 | 19,891 |
| #8 | Treatment of 5 Critically Ill Patients With COVID-19 With Convalescent Plasma | 10.1001/jama.2020.4783 | JAMA | 2020-03-27 | 19,152 |
| #9 | Covid-19 - Navigating the uncharted. | 10.1056/nejme2002387 | New England Journal of Medicine | 2020-03-26 | 18,096 |
| #10 | Do us a favor. | 10.1126/science.abb6502 | Science | 2020-03-13 | 18,038 |

The Twitter mentions of the top-10 publications were depicted in Figure 2 based on their post date. Twitter mentions of different publications are highlighted by different colors in each bar according to their proportion in total Twitter mentions posted in that day. The publication date of each publication is annotated on the top of the corresponding bar, to show how fast the Twitter attention was accumulated after they were published. In general, Twitter attention toward a specific highly tweeted publication concentrated within the first few days after its publication. For example, publication #4, as a preprint ranking in the top ten which has been withdrawn by the authors, attracted a high level of Twitter attention in the very early stage of the coronavirus crisis compared to other publications. Most of the Twitter mentions of publication #4 occurred in the first two days after its publication. Similar patterns can be observed for most other highly tweeted publications. Moreover, due to the existence of advanced online version, some publications have already accumulated a certain amount of Twitter attention before their publication date recorded (e.g., publication #9 and #10). In a word, in the case of the top-10 most tweeted publications, Twitter users were following COVID-19-related research in next to no time since their appearance, besides, the Twitter attention concentrated in the first few days and then gradually faded away. This finding is in line with what has been observed in a previous study based on a more generalized dataset (Shuai et al., 2012), in which Twitter mentions were also found to have relatively short delays and narrow time spans.

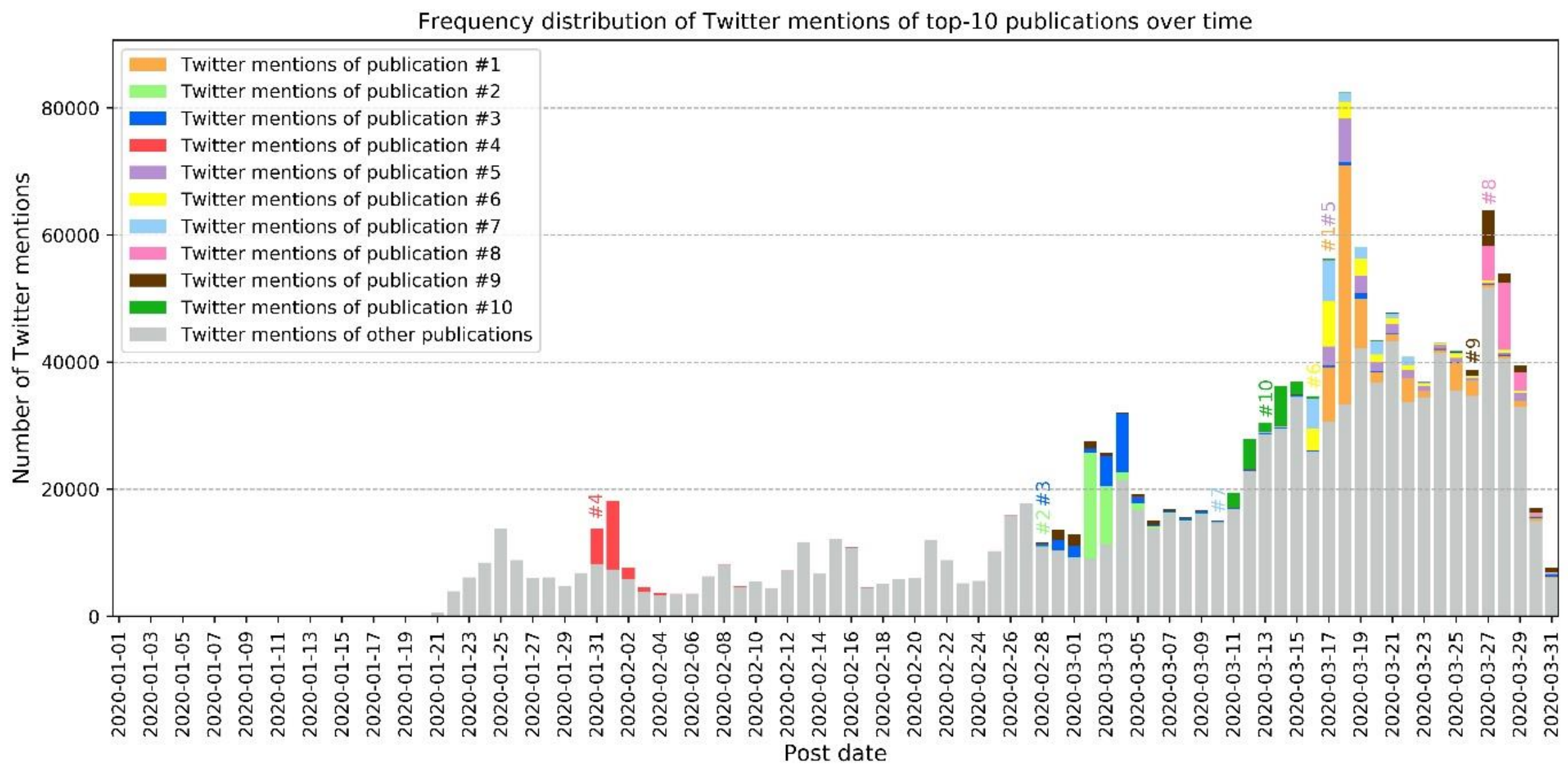

**Figure 2**. Distribution of tweets of the top-10 most tweeted publications posted in different date.

*Characterizing the Twitter users sharing COVID-19-related research*

In our dataset, there are 161,390 unique users who have posted at least one tweet with original contents about COVID-19-related publications, by whom the research outputs were initially brought into the Twitter landscape. To reveal the characteristics of the users, by drawing on their profile descriptions written in English, Figure 3(a) illustrates the co-occurrence network of terms that users used to describe themselves. Size of nodes is proportional to their frequency of occurrence. For clear visualization, the 269 most relevant terms were kept, thereby algorithmically generating four clusters. Cluster #1 (red cluster) mainly contains terms that belong to *personal description* about the private and personal life, like “husband”, “father”, and “mom”, etc. Cluster #2 (blue cluster) is formed by terms implying users’ *interest and faith*, like “music”, “dog”, and “Christian”, etc. Cluster #3 (green cluster) is comprised by terms referring to *academic role* in the scientific community, such as “professor”. “PhD student”, and “university”, etc. At last, cluster #4 (yellow cluster) includes terms that express users’ *attitude and position* on Twitter, like “endorsement”, “like”, and “retweet”, etc. In general, the clustering is consistent to a large extent with the main textual patterns of Twitter users’ descriptions found in the study conducted by Díaz-Faes et al. (2019), which is based on a more universal dataset consisting of a random sample of 200,000 Twitter users who have ever tweeted at least one scholarly output. According to their results, terms extracted from profile descriptions are clustered into four communities as well, including personal description, academic role, business and practice role, and attitude and position. The difference of involved Twitter users between the universal situation and the COVID-19 case is that users with business and practical roles are less active, while those highlighting their interest and faith and academic roles become more predominant in the case of COVID-19, indicating that research progress taken place on COVID-19 might attracted a higher proportion of general users and scientific researchers.

After summing up the number of retweets received by all tweets with original contents, each Twitter user considered in Figure 3(a) were assigned with their own total number of obtained retweets. As a result, in Figure 3(b) terms were scored by the average number of retweets that Twitter users have received. Terms related to academic roles have predominately higher retweet scores, suggesting that Twitter users describing themselves with academic roles are more likely to get retweeted in the case of COVID-19, especially for senior researchers and users from the biomedical and health sciences-related fields such as epidemiology, immunology, and cardiology. Therefore, during the COVID-19 crisis, scholarly-oriented Twitter users with expertise seem to be playing an influential role in updating and disseminating science-based information on Twitter.

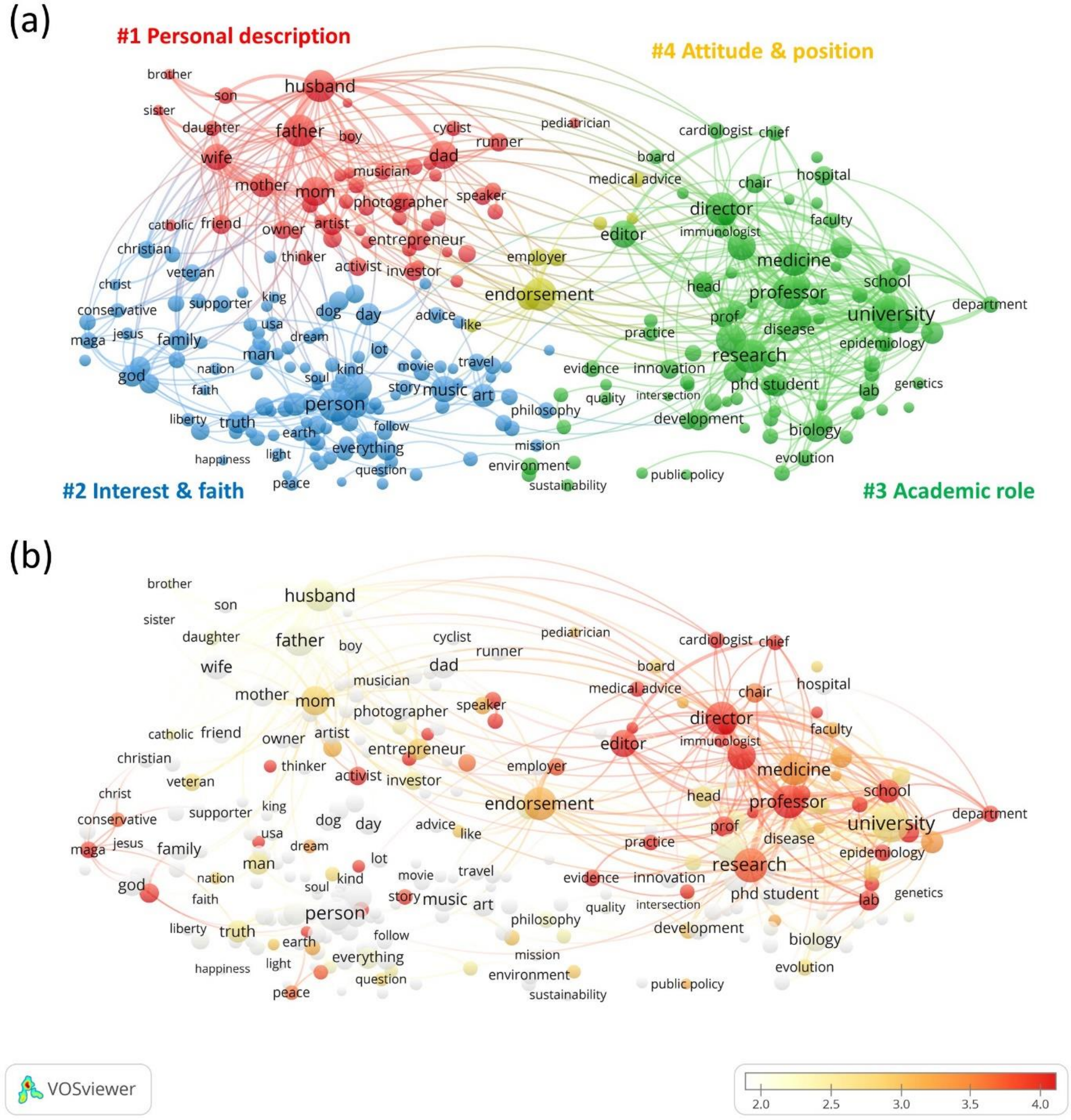

**Figure 3**. (a) Co-occurrence network of terms used in the profile descriptions of Twitter users who have posted tweets with original contents related to COVID-19-related publications. (b) Overlay visualization of terms scored by the average number of retweets that Twitter users have received.

*Identifying topics of COVID-19-related research mentioned on Twitter*

In order to show an overview of the research topics that authors focused on, Figure 4(a) presents the co-occurrence network of terms used in the titles of 6,162 COVID-19-related publications. Size of nodes is determined by their frequency of being used. For the sake of clear visualization, we exhibited the 145 most relevant terms, which are aggregated into three clusters. The three clusters generated correspond to three main research directions on COVID-19: cluster #1 (blue cluster) includes terms related to *epidemiological research*, with special attention to the coronavirus pandemic in different countries and their countermeasures; cluster #2 (red cluster) mainly focuses on *virological research*, especially for the protein structure of the virus and its proximal origin; and cluster #3 (green cluster) is comprised of terms about *clinical research*, and in this direction, most terms are associate with clinical characteristics and coronavirus infection.

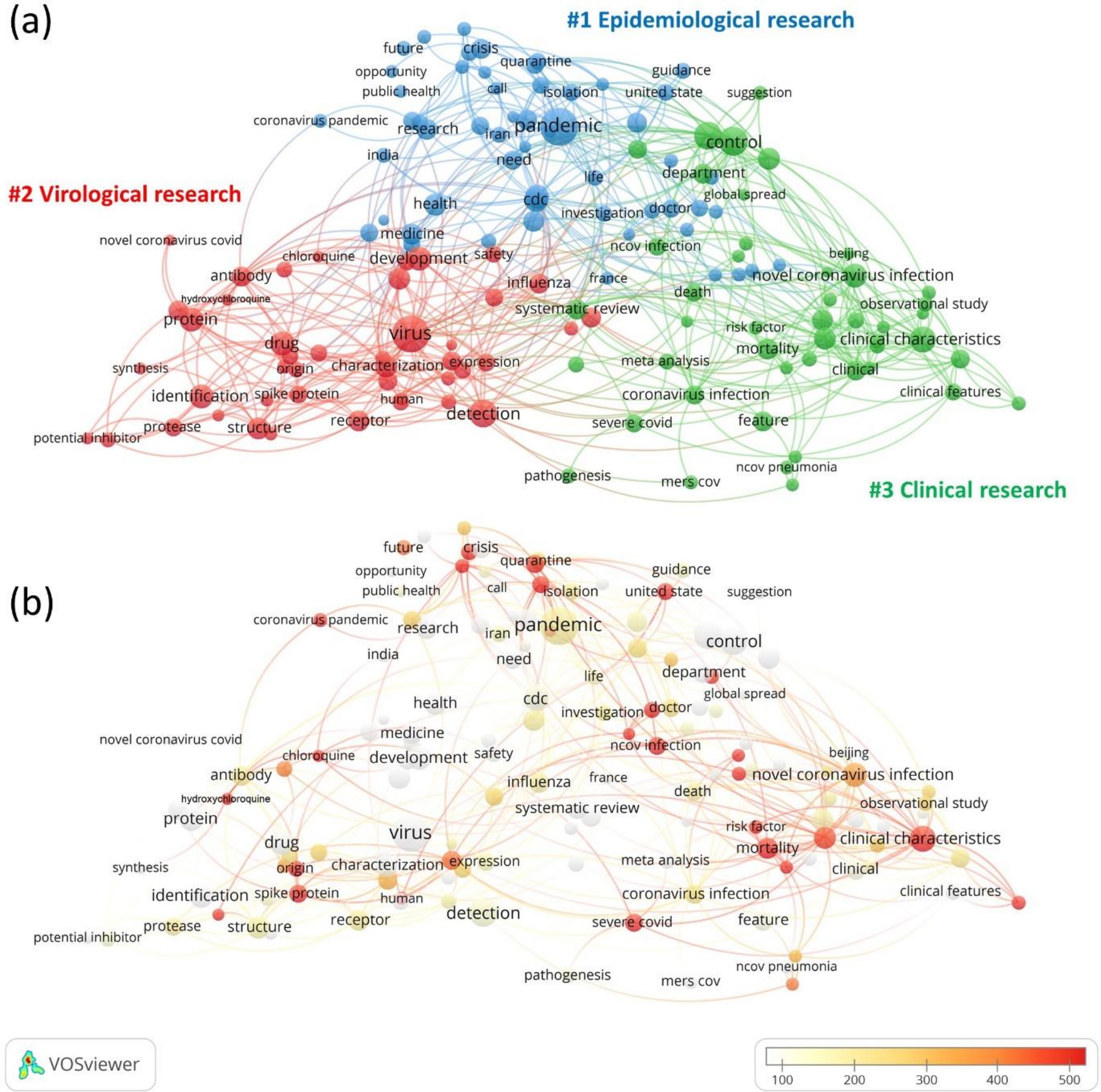

**Figure 4**. (a) Co-occurrence network of terms used in the titles of COVID-19-related publications. (b) Overlay visualization of terms scored by their average number of Twitter mentions received.

On the basis of the total number of Twitter mentions that each publication has accumulated, terms were scored by the average number of Twitter mentions accrued, as shown in Figure 4(b). As to each research direction, there exist some research topics with relatively intensive Twitter attention. For example, in the field of epidemiology, the actions taken by the governments against the coronavirus pandemic, which have made a significant impact on people's daily lives, are of interest to Twitter users, like "quarantine" and "lockdown". In the field of virology, Twitter users paid quite a lot of attention to the exploration of the animal source of the novel virus as scientists did (Mallapaty, 2020), making "origin" and related genomic structure terms the most tweeted topics in this field. In addition, "chloroquine" and "hydroxychloroquine", the two widely-used anti-malarial drugs, attracted a great deal of attention as well, because of their controversial reported efficacy and safety in inhibiting the novel coronavirus in vitro and in different clinical trials (Gautret et al., 2020; Liu et al., 2020; Molina et al., 2020; Wang et al., 2020). However, on the whole, topics in clinical research got higher levels of Twitter attention. As the topics directly related to the severity of the pandemic and the health of humans, "clinical characteristics", "mortality", "ncov infection", "risk factor", and some other research topics have been of great concern by the public.

While sharing COVID-19-related research on Twitter, users might express their own opinions on or beyond the tweeted research. To quantitatively reflect the extent to which science-based Twitter conversations are different from the titles of publications, Figure 5 shows the results of the calculation of cosine similarity scores between publication titles and related full tweet texts. There are around 45.3% of tweets with original contents added showing a big dissimilarity with the mentioned publications' titles, with the cosine similarity scores falling between 0 and 0.1. Over 74% of Twitter mentions have the values of similarity lower than 0.5, implying that most Twitter users conducted some commenting rather than just repeating the titles of the publications. This finding is further confirmed by the gap between the average length of full tweet texts and that of publication titles in each statistical bin of similarity scores. Most Twitter mentions are formed by longer texts than titles, indicating the extension of contents made by Twitter users while sharing scientific information. Different from the devoid of original thought of tweet contents observed in previous studies based on the samples from the fields of dentistry and big data (Lyu & Costas, 2020; Robinson-Garcia et al., 2017), during the COVID-19 crisis, Twitter users were more inclined to set forth their original opinions and attitudes when tweeting about recent scientific outcomes.

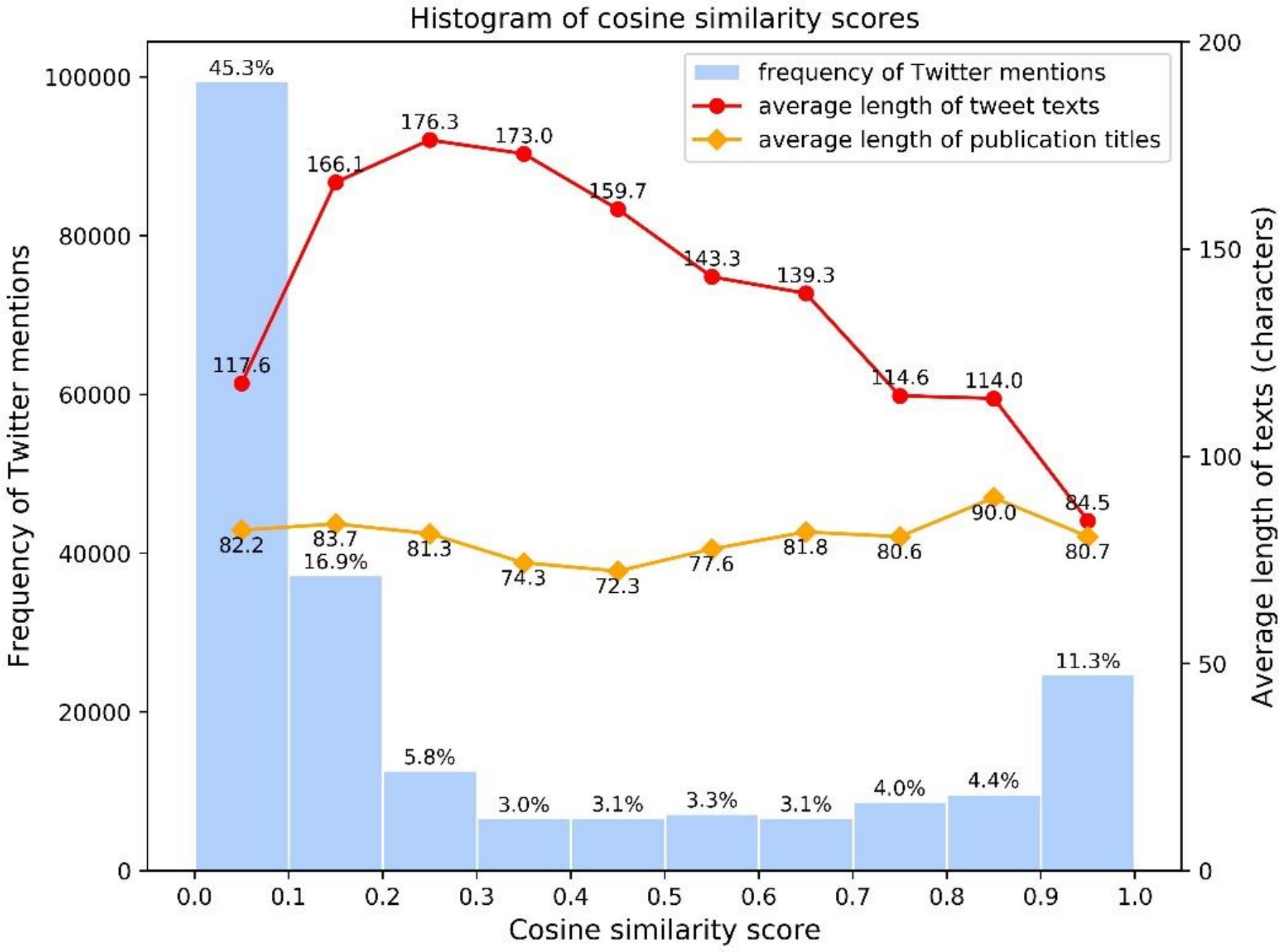

**Figure 5**. Frequency distribution of Twitter mentions with different cosine similarity scores between the titles of publications and tweet texts, as well as the average length of publication titles and tweet texts in each statistical bin.

*Tracking the Twitter discussions around COVID-19-related research over time*

Given that Twitter users generally enlarged the contents of their tweets beyond the publication titles, in this part we further explored what they talked about around COVID-19-related research and how the focus of their discussions changed over time. Figure 6(a) shows the co-occurrence network of terms used in the tweets with original contents in English. Size of nodes is proportional to the occurrence frequency of each term. As the corpus of full tweet texts is much larger than that of publication titles, we kept more terms in the map to present a relatively complete landscape of science-based Twitter conversations. The 601 most relevant terms were retained for clear visualization, thus generating five clusters of the main topics in the Twitter discussions related to COVID-19-related research.

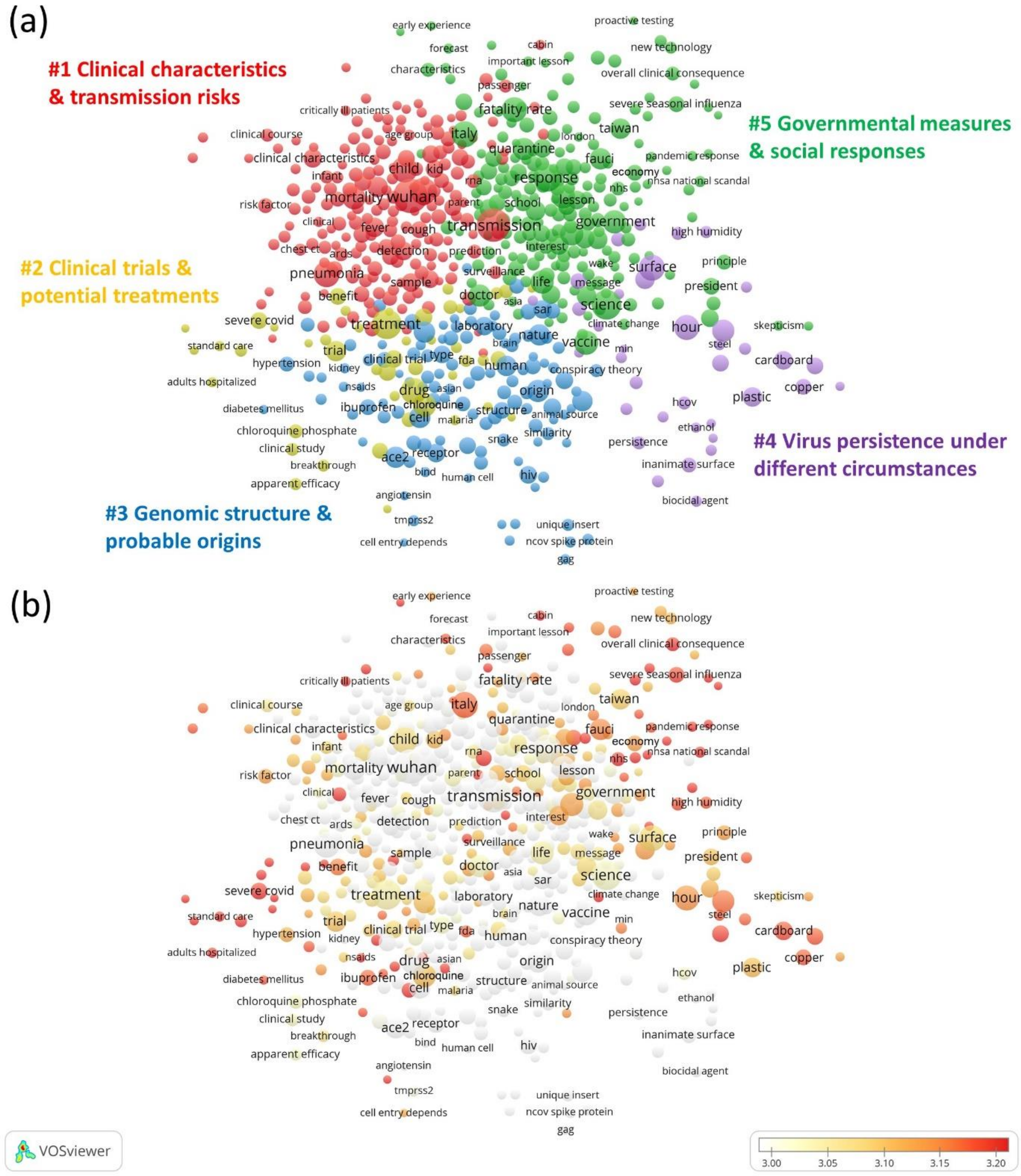

**Figure 6**. (a) Co-occurrence network of terms used in the tweets with original contents in English; (b) Overlay visualization of terms scored by the average post date of tweets in which they are mentioned.

Terms in cluster #1 (red cluster) reveal the public concerns about the *clinical characteristics and transmission risks* of the infectious disease caused by the virus. Besides some specific symptoms (e.g.,

"fever" and "cough") and the "mortality", the risk of being infected was also seriously concerned for different age groups, particularly for "child/kid". These terms were frequently occurred with "Wuhan", the initial city struggled with the novel coronavirus. Terms in cluster #2 (yellow cluster) reflect the discussions about *clinical trials and potential treatments* that have been conducted to treat or prevent COVID-19, including some general terms like "treatment", "trial", and "drug", etc., as well as more specific therapies like "chloroquine" and "hydroxychloroquine", the two controversial drugs as aforementioned which have caused a wave of misinformation due to the unproven effectiveness in treating COVID-19 (Erku et al., 2020). As the basis of knowing more details of the causative coronavirus, terms about its *genomic structure and probable origins* were also frequently mentioned by Twitter users in cluster #3 (blue cluster), including not only the jargons of genomics (e.g., "receptor" and "ACE2") but also the comparisons of similarity with other viruses like "HIV", based on which Twitter users discussed about several animal sources supposed to be the probable origins of the coronavirus in the nature. Cluster #4 (purple cluster) includes a sequence of terms about the *virus persistence under different circumstances*, especially on various surfaces (e.g., "cardboard", "plastic", and "copper") and in "high temperature" and "high humidity". Last but not least, terms related to the *governmental measures and social responses* constitute cluster #5 (green cluster), which contains a range of topics reflecting the tremendous influences that COVID-19 have exerted on human society, such as the countermeasures applied by the governments across countries and regions, and the related issues emerged in education (e.g., "school") and "economy".

To track the evolution of the focus of Twitter discussions, as shown in Figure 6(b), we assigned each term with the average post date score. In general, terms in cluster #1 and #3 that related to the clinical characteristics and genomic structure are in lighter colors because they were mentioned by Twitter users in the relatively earlier stage of the coronavirus outbreak. In contrast, terms in cluster #2, #4, and #5 are in darker colors, suggesting that with the spread of the COVID-19 pandemic, the focus of the Twitter discussions transferred from clinical and virological research to topics on how to treat and prevent the infectious disease, how do the governments face the public health emergency, and how does the pandemic affects society.

Considering that hashtags are deemed as *concept symbols* indicating particular concepts in relation to the mentioned publications (Haustein et al., 2016), just like author keywords to scientific publications (Haunschild et al., 2019), hereby Figure 7(a) shows the hashtag coupling network of 219 English hashtags that have been used at least 50 times in total.[8] There are four clusters generated based on these hashtags: hashtags in cluster #1 (green cluster) are mainly about the coronavirus outbreak in China and virological research; cluster #2 (yellow cluster) includes hashtags of some medical technologies and hashtags labelling the PubMed updates; hashtags in cluster #3 (red cluster) highlights some political and misinformation debates, in which "chloroquine" and "hydroxychloroquine" are included; finally, hashtags in cluster #4 (blue cluster) indicates the global spread of the outbreaks in many countries and a series of healthcare measures that were called on to prevent the virus transmission.

Similarly, Figure 7(b) shows the overlay visualization of the hashtag coupling network scored by the post date. It is obvious that hashtags located in the lower right part hold higher post date scores than those in the upper left part, which means that in the early stage of the coronavirus outbreak, the pandemic in China was frequently discussed by Twitter users, as well as the basic virological research on the novel virus. However, along with the global spread, the focus of Twitter attention transferred to other countries suffering from the crisis, by the meantime, Twitter users started to attach hashtags of healthcare measures, such as "#flattenthecurve", "#socialdistancing", and "#stayhome", to underline the importance of taking actions to stop the virus.

[8] For clear visualization, we excluded the dominant hashtags that only contain the names of the infectious disease or the novel coronavirus or their variants from the network, such as "#COVID19", "#COVID_19", and "#SARS_COV_2".

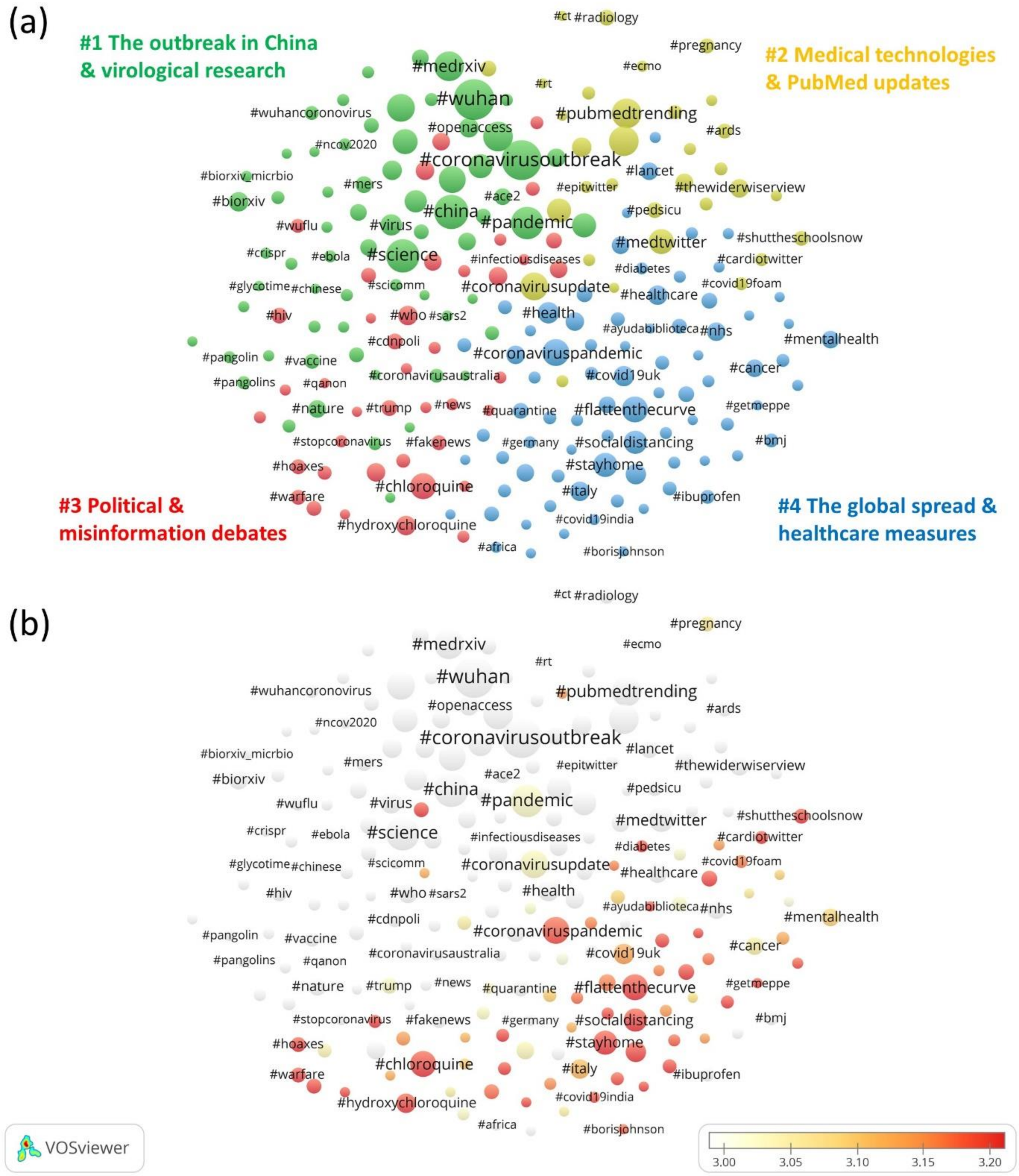

**Figure 7**. (a) Coupling network of English hashtags; (b) Overlay visualization of hashtags scored by the average post date of tweets in which they are attached.

## Discussion

### *The tweeting patterns around COVID-19-related research*

The global outbreak of COVID-19 has triggered an avalanche of scientific research and public discussions, thereby generating a so-called *infodemic* referring to an over-abundance of information related to the epidemic that increases the risk of misinformation,[9] which is also concerned by Twitter users in the light of their use of hashtags pointing controversial misinformation. As emphasized by Xie et al. (2020), it is of great importance to study information behaviors during global health crises. Against this background, we focused on the information behaviors of sharing and discussing COVID-19-related research on Twitter. By employing nearly 1.4 million Twitter mentions as the traces of broader public engagement with COVID-19-related publications, we tracked the changing public concerns about the novel PHEIC during the first three months of 2020.

Since the initial outbreak of COVID-19 at the very end of 2019 in China, there have been a growing torrent of new scientific publications on this infectious disease and the causative novel coronavirus (Brainard, 2020). As one of the most significant windows for the public to get better understanding of the unprecedented pandemic, continuous research progress has widely attracted Twitter attention from the public, causing a wealth of Twitter mention data that connect the science landscape and the social media landscape. For the 6,162 COVID-19-related publications in our dataset, 68.1% of them have been mentioned on Twitter at least once. The value of Twitter coverage is much higher than those found in previous surveys based on larger-scale multidisciplinary samples (13.3% by Costas et al. (2015), 21.5% by Haustein et al. (2015)), implying that the Twitter attention to COVID-19-related research is much larger than usual. Alongside the pandemic of COVID-19 entered a new stage with rapid spread in countries outside of China in the middle of March, 2020 (Bedford et al., 2020), the Twitter attention toward research progress increased drastically in late March accordingly.

Overall, the tweeting patterns around COVID-19-related research comply with the observed general tweeting patterns from the perspectives of both temporal accumulation and involved Twitter users. Through the lens of the top-10 most tweeted publications, we found that Twitter attention accumulated soon after the publication and concentrated in the following first few days, which is in accordance with the general temporal accumulation patterns of scholarly Twitter mentions (Fang & Costas, 2020; Shuai et al., 2012). Based on the users' profile descriptions, the composition of Twitter users with interest in discussing COVID-19-related research was found to be similar with the general communities of Twitter users who shared scholarly literature (Díaz-Faes et al., 2019). Twitter users primarily described themselves from the angles of personal description, interest and faith, academic roles, and attitude and position. In terms of being retweeted, Twitter mentions originated from users with academic roles got substantially more retweets, especially for senior researchers and users describing themselves as being related to the field of biomedical and health sciences. Therefore, in the case of COVID-19, science-based information posted by academically related users are more influential in the Twitter dissemination network.

*The evolution of the focus of Twitter discussions over time*

Based on our dataset, topics extracted from the titles of COVID-19-related publications were clustered into three main research directions, including epidemiological research, virological research, and clinical research. In each direction, there are some research topics with higher levels of Twitter attention received. Same as what has been observed by Colavizza et al. (2020), in general, research topics of pressing relevance received more Twitter attention, especially for those in relation to clinical features, infection, treatments, and countermeasures.

Different from the strong concordance between scholarly contents and related tweet contents found in other fields (Lyu & Costas, 2020; Robinson-Garcia et al., 2017), the Twitter discussions around COVID-19-related research show a higher degree of engagement according to the dissimilarity between publication titles and related full tweet texts. Instead of simply repeating the titles, Twitter users expanded the tweet contents with their own opinions in the science-based Twitter conversations. The expanded Twitter discussions are mainly about five topics related to the COVID-19 pandemic, including the clinical characteristics and transmission risks, the clinical trials and potential treatments, the genomic structure and probable origins, the virus persistence under different circumstances, and the governmental measures and social responses. On the one hand, these topics frequently mentioned by Twitter users have a strong connection to the scientific discoveries; on the other hand, Twitter users talked about COVID-19-related research with particular focus on its impacts on the real world, covering a range of concerns about children, education, economy, and society.

[9] The concept of the term "infodemic" was mentioned in the WHO Situation Report-13 at: https://www.who.int/docs/default-source/coronaviruse/situation-reports/20200202-sitrep-13-ncov-v3.pdf?sfvrsn=195f4010_6. Accessed 2020-05-20.

Besides, there is a change trend of the focus of the Twitter discussions across the above topics over time. At the early stage when the COVID-19 pandemic was confined to China, most Twitter discussions were associated with the research on the virological and clinical characteristics, however, over time the focus of the Twitter discussions transferred to the pursuit of effective treatments, the countermeasures by the governments, and the strong influences on social and economic activities. This finding was further confirmed by the usage of hashtags, with the hashtags related to China and basic virological research dominated at the beginning and those about global outbreaks, potential treatments, and healthcare measures stood out latterly. In light of these results, during the age of worldwide pandemic, it appears that understanding the origin and features from the virological and clinical aspects was the first priority for the public concerns before a global outbreak, while along with the aggravation of situation and the improvement of cognition, the public concerns transferred to a broader scope beyond academia, particularly about the progress in treatment, the countermeasures by the governments, the healthcare measures for self-protection, and the influences on society.

The long-term impacts made by COVID-19 are still continuing, in future research, we will continue to study the evolution of the Twitter attention toward COVID-19-related research. To unveil the differences of local public concerns, we plan to respectively explore the Twitter attention from different geographic locations and in different language contexts.

*Limitations*

There are some limitations that should be acknowledged in this study. First, we relied on the bibliometric information provided by Dimensions.ai for most analyses, so only COVID-19-related publications with DOI or PubMed ID indexed by Dimensions.ai were taken into account, while those without available identifiers or not indexed by Dimensions.ai were not included in our analyses. Second, due to the limitation of bibliometric information that we retrieved from Dimensions, only the titles were analyzed as the representatives of scholarly contents of COVID-19-related publications, while other bibliometric information like abstract, keywords, and full texts were left out. Lastly, in the analyses of full tweet texts and hashtags, we only considered the tweets written in English. Although English tweets account for the majority of the Twitter discussions, they might only reflect the opinions owned by English-speaking users.

## Conclusions

This study tracked the Twitter attention surrounding the COVID-19-related research efforts during the first three months of 2020, to disclose the public concerns about the COVID-19 pandemic embedded in the science-based conversations on Twitter. Twitter mentions of COVID-19-related publications presented a rising tendency in parallel with the increasing number of research outputs, especially in late March, 2020. Amongst the Twitter users sharing COVID-19-related research, academically related users with relevant expertise played an influential role in disseminating research outputs on Twitter according to their higher levels of retweets received. In general, research topics with pressing relevance attracted more Twitter attention, such as those about clinical characteristics, infection, treatments, and countermeasures. With time elapsing, the focus of the Twitter discussions around COVID-19-related research evolved from virological and clinical research findings to the potential treatments, the countermeasures by the governments, the healthcare measures for self-protection, and some broader discussions about the social influences, showing a dynamic of the public concerns during the age of public health emergency. Overall, our results support the idea that the analysis of the Twitter engagement around the COVID-19-related research efforts can provide evidence of other types of societal perceptions around the pandemic. Therefore, the study of the social media response to the scholarly response, can be a useful approach to inform the existence and development of broader concerns around the pandemic.

## Acknowledgements

Zhichao Fang is financially supported by the China Scholarship Council (201706060201). Rodrigo Costas is partially funded by the South African DST-NRF Centre of Excellence in Scientometrics and Science, Technology and Innovation Policy (SciSTIP). The authors thank Prof. Paul Wouters (Leiden University) for helpful suggestions, and Giovanni Colavizza (University of Amsterdam) and Nees Jan van Eck (Leiden University) for collecting and parsing the bibliographic meta-data. The authors also thank Dimensions.ai and the WHO for compiling and opening COVID-19 research data, and Altmetric.com and Twitter for providing APIs to access the Twitter mention data.